\documentclass[twocolumn,superscriptaddress,aps,preprintnumbers,amsmath,amssymb,sort&compress,nofootinbib,prl]{revtex4-1}

\usepackage{amsfonts,amsmath,amssymb,bm}
\usepackage{comment}
\usepackage{ifpdf}
\usepackage{slashed}
\usepackage{color}
\usepackage[mathscr]{eucal}
\usepackage[utf8]{inputenc}
 
\usepackage{cancel}

\ifpdf
  \usepackage{graphicx}     %   usepackage without driver option
  \usepackage[bookmarksopen,colorlinks=true,linkcolor=bblue,citecolor=bblue,urlcolor=ppink]{hyperref}
\else     % For (p)LaTeX + dvipdfmx
  \usepackage[dvipdfmx]{graphicx}     %   usepackage with driver option
  \usepackage[dvipdfmx,bookmarksopen,colorlinks=true,linkcolor=bblue,citecolor=ppink,urlcolor=darkred]{hyperref}
\fi

\definecolor{red}{rgb}{1,0,0}
\definecolor{darkred}{rgb}{0.6,0,0}
\definecolor{darkgreen}{rgb}{0.992447,0.623778,0.034597}
\definecolor{ppink}{rgb}{1,0.4,0.4}
\definecolor{bblue}{rgb}{0.284602,0.317763,0.963947}

\makeatletter
\newcommand\footnoteref[1]{\protected@xdef\@thefnmark{\ref{#1}}\@footnotemark}
\makeatother

\begin{document}

%%%%%%%%%%%%%%%%%%%%%%%%%%%
%%%%%%%%%%% Title %%%%%%%%%%%
%%%%%%%%%%%%%%%%%%%%%%%%%%%

%%paper
\title{Gravitational waves, CMB polarization, and the Hubble
tension}
\author{Donghui Jeong}
\affiliation{Department of Astronomy and Astrophysics and Institute for Gravitation and the Cosmos, \\
The Pennsylvania State University, University Park, PA 16802, USA
}\author{Marc Kamionkowski}
\affiliation{Department of Physics and Astronomy, Johns Hopkins University, 3400 N. Charles Street,
Baltimore, MD 21218, U.S.A.}

\begin{abstract}
The discrepancy between the Hubble parameter inferred from local 
measurements and that from the cosmic microwave background (CMB)
has motivated careful scrutiny of the assumptions that enter
both analyses.  Here we point out that the location of the recombination 
peak in the CMB B-mode power spectrum is determined by the light horizon 
at the surface of last scatter and thus provides an alternative 
early-Universe standard ruler. It can thus be used as a
cross-check for the standard ruler inferred from the acoustic peaks 
in the CMB temperature power spectrum and to test various explanations 
for the Hubble tension. The measurement can potentially be carried out
with a precision of $\lesssim2\%$ with stage-IV B-mode experiments. 
The measurement can also be used to measure the propagation speed of gravitational waves in the early Universe.
\end{abstract}

\date{\today}
\maketitle
%\preprint{}

The tension \cite{Freedman:2017yms,Feeney:2017sgx,Verde:2019ivm}
between the value of the Hubble parameter (the cosmic expansion rate)
inferred from local measurements
\cite{Riess:2016jrr,Riess:2018byc,Bonvin:2016crt,Birrer:2018vtm}
and that \cite{Ade:2015xua,Aghanim:2018eyx} inferred from the
cosmic microwave background (CMB) has been lingering for a
number of years.  It is now established at the $\gtrsim 4\sigma$
and should rightfully be promoted from a Hubble ``tension'' to a
{\it bona fide} discrepancy.  The discrepancy is not easily
attributed to any obvious systematic error
\cite{Efstathiou:2013via,Addison:2015wyg,Aghanim:2016sns,Aylor:2018drw}.
Several recent papers have shown that the local
measurements, which are obtained by comparing the inferred
distance to cosmological sources with their redshifts,
are robust to new or alternative calibrations of the cosmic distance ladder
\cite{Riess:2019cxk,Pietrzynski:2019,Taubenberger:2019qna,Collett:2019hrr}.
Note, however, the recent debate \cite{Freedman:2019jwv,Yuan:2019npk} 
with the calibration using the TRGB stars.
The most recent local measurement is $H_0=74.22\pm 1.82$
km/sec/Mpc \cite{Riess:2019cxk}.  On the other hand, the Hubble
parameter is inferred from the CMB from the angular scale of
peaks in the CMB  angular power spectrum.  This angular scale is
fixed by the ratio of the sound horizon 
(the distance a sound wave in the primordial baryon-photon fluid has 
traveled from big bang to the time the CMB decoupled) 
with the angular-diameter distance to the surface of last 
scatter \cite{Kamionkowski:1993aw,Jungman:1995av}. Both distances 
are obtained, within the standard cosmological model, by detailed
modeling of the CMB peak structure. This procedure yields a
value $H_0=67.4\pm0.5$ km/sec/Mpc \cite{Aghanim:2018eyx}.

Solutions to the Hubble tension are not easily come by but
generally involve modifications to cosmic evolution at early
times (mechanisms that decrease the sound horizon)
\cite{Karwal:2016vyq,Riess:2016jrr,Riess:2018byc,Bernal:2016gxb,Poulin:2018cxd,Lin:2018nxe,Kreisch:2019yzn,Blinov:2019gcj,Park:2019ibn,Agrawal:2019lmo,Lin:2019qug}
or at late times (modifications to the cosmic expansion history that
increase the angular-diameter distance to the surface of last scatter)
\cite{DiValentino:2016hlg,DiValentino:2017zyq,DiValentino:2017rcr,DiValentino:2017iww,Pandey:2019plg,Vattis:2019efj}. However,
the late-time resolutions are tightly constrained by other late-time observables
\cite{Riess:2016jrr,DiValentino:2017zyq,Addison:2017fdm,DiValentino:2017iww,Beutler:2011hx,Ross:2014qpa,Alam:2016hwk,Bernal:2016gxb,Zhao:2017cud,Poulin:2018zxs},
and the early-time solutions are tightly constrained by the
acoustic oscillations in the CMB power spectrum. All of the
proposed solutions require fairly exotic new physics.

Given the lack of any easy solutions to the Hubble tension, as
well as the increasing significance of the discrepancy,
any possible cross-checks of the measurements and assumptions,
as well as any possible complementary information that can be
obtained, should be pursued vigorously. In particular, all the
information we have about the Hubble parameter relies ultimately
on distance measures in cosmology, and any new technique to
obtain a cosmic distance will be valuable.

We propose that measurement of the B-mode polarization in the
CMB \cite{Kamionkowski:1996ks,Zaldarriaga:1996xe} induced by
primordial gravitational waves
\cite{Kamionkowski:1996zd,Seljak:1996gy,Kamionkowski:2015yta} may be used to
provide an independent cross-check of the early-Universe
expansion history.  These B modes have yet to be detected but
are predicted in the canonical single-field slow-roll inflation
models to be within the sensitivities of major experimental
efforts---for example, 
CLASS \cite{Essinger-Hileman:2014pja},
LiteBIRD \cite{Matsumura:2013aja}, 
the Simons Observatory \cite{Ade:2018sbj},
CMB-S4 \cite{Abazajian:2016yjj},
or Probe Inflation and Cosmic Origins (PICO
\cite{Hanany:2019lle})---to be pursued within the
next decade. If they exist and are detected, they may prove to
be of value in efforts to understand the Hubble tension.

The primordial B-mode power spectrum exhibits oscillations that arise from
the propagation of gravitational waves
\cite{Pritchard:2004qp,Flauger:2007es}.  These are analogous to
the well-known acoustic oscillations in the CMB temperature power
spectrum \cite{Sunyaev:1970eu,Peebles:1970ag} that arise from sound
waves in the primordial baryon-photon fluid.  The
difference is that the propagation speed of sound waves in the
photon-baryon fluid is roughly $c/\sqrt{3}$, while that of
gravitational waves is the speed of light $c$.

If the Hubble
tension is due to a late-time modification of the expansion
history, both sets of peaks (those in the temperature power
spectrum and those in the GW-induced B-mode power spectrum)
should be affected in the same way.  The peaks in the B-mode
power spectrum should thus appear at the same multipole moment
as predicted in the current
best-fit cosmological model.  If the discrepancy is resolved by
new physics in the
early Universe, the peak locations in the B-mode power spectrum
may differ.  More precisely, the comoving sound horizon at decoupling is an
integral $r_s = \int^{t_{\rm ls}}\, c_s(t) dt/a(t)$ of the sound
speed $c_s(t)$ until the time $t_{\rm ls}$ of CMB decoupling, while the
comoving gravitational-wave horizon is $r_{\rm gw} =
c\int^{t_{\rm  ls}}\, dt/a(t)$.  If the Hubble tension is resolved
somehow by a reduction in the sound speed, then the B-mode peak location,
relative to the acoustic peak, will change.  Existing models
generally involve some shift in the expansion history (which
affects $r_s$ and $r_{\rm gw}$ in a slightly different way) and some shift in the
baryon and dark-matter densities (which can affect the two
distances differently).

\begin{figure}[t]
    \includegraphics[width=0.49\textwidth]{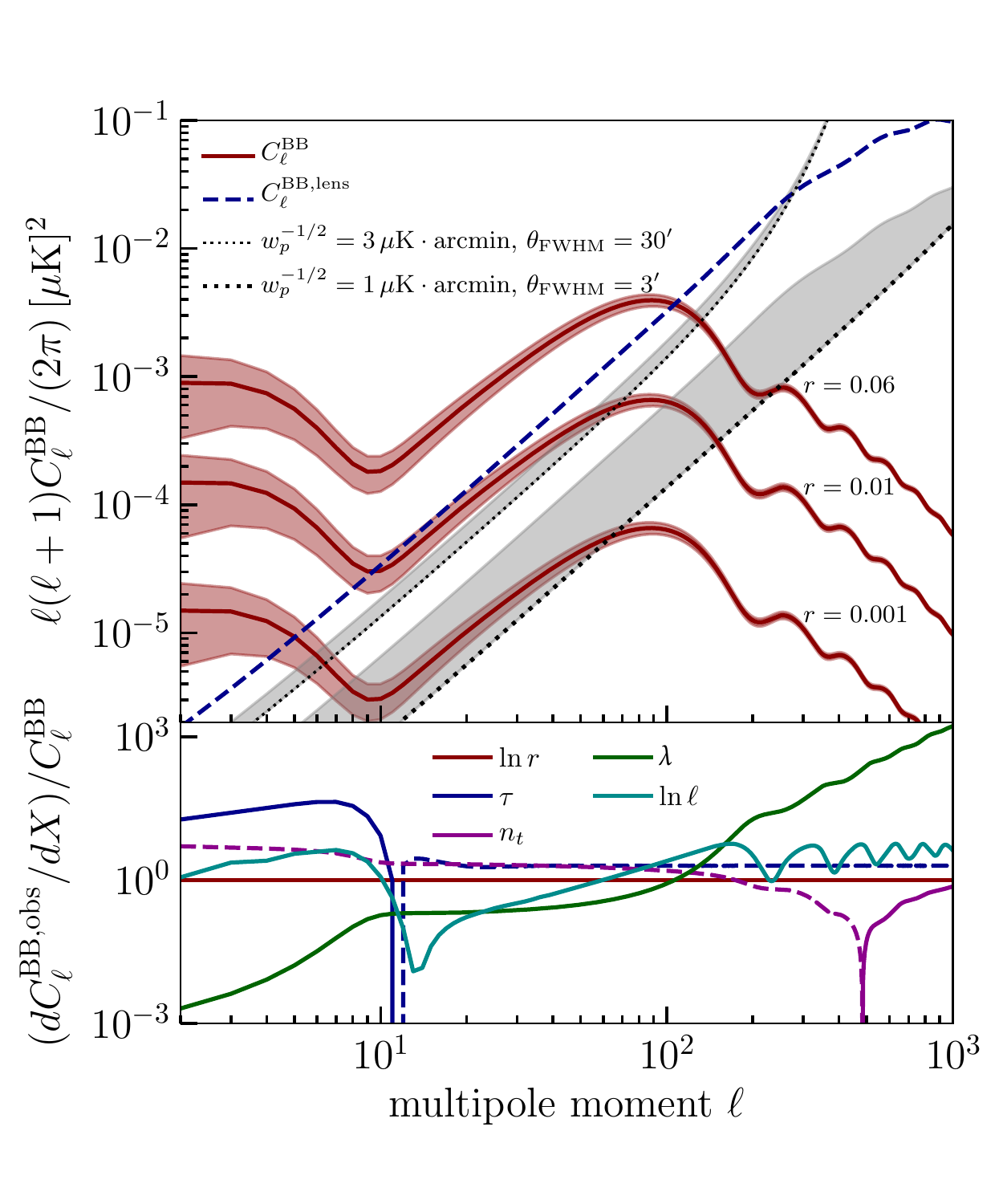}
\caption{
{\it Top}: The B-mode polarization power spectrum ({\it red lines}) for 
$r=0.06$, $0.01$ and $0.001$ (from top to bottom), along with
the cosmic-variance uncertainty ({\it red shaded regions}) and
instrumental noise 
({\it black lines}) similar to LiteBIRD ({\it dotted}) and CMB stage-IV 
or PICO ({\it dashed}). The gray shaded regions show the extra noise
contribution from the gravitationally lensed B-mode ({\it blue line}) 
between perfect removal case (bottom edge) and $15\%$ delensing residual
(upper edge).
{\it Bottom}: 
The response of the B-mode polarization power spectrum $d\ln C_\ell/d X$ to the change of parameters: $X= \ln r$ (tensor amplitude), 
$\tau$ (optical depth), $n_t$ (tensor spectra index), 
$\lambda$ (gravitational lensing amplitude), and 
$\ln \ell$ (distance scales). For the Fisher analysis, we set 
$d C_\ell/d\ln\ell=0$ for $\ell<15$ in order to exclude the reionization 
bump.}\label{fig:ClBB}
\end{figure}

To be relevant for the $\Delta H_0/H_0\sim10\%$ tension, 
the angular scale of the peaks in the B-mode power spectrum must be
determined to better than 10\% (the magnitude of the discrepancy).
As the calculation below indicates, this is conceivable with measurements 
to be carried out on a decade timescale. The measurement is, however, by 
no means guaranteed, even if the experiments perform as expected, as
the determination requires that primordial gravitational waves
(which are hypothesized but have yet to be detected) have an
amplitude $r \gtrsim 0.001$ (see Fig.~\ref{fig:ClBB}).  Here,
$r$ is the tensor-to-scalar ratio of the primordial power
spectra.

%%%%%%%%%%%%%%%%%%%%%%%%%%%%%%%%%%%%%%%%%%%%%%%%%%%%%%%%%%%%%%%%%%%%%%%%%%
%%%%%%%%%%%%%%%%%%%%%%%%%%%%%%%%%%%%%%%%%%%%%%%%%%%%%%%%%%%%%%%%%%%%%%%%%%

In this paper, we study the possibility to determine the light-horizon 
scale from future B-mode polarization experiments such as 
LiteBIRD \cite{Matsumura:2013aja}, 
a CMB stage-IV experiment (e.g., the Simons Observatory
\cite{Ade:2018sbj} or CMB-S4 \cite{Abazajian:2016yjj}), or
Probe Inflation and Cosmic Origins (PICO \cite{Hanany:2019lle}). 
These efforts aim to detect the primordial B-mode polarization
with sentivity better than $\sigma_r\sim0.001$.

We first begin with some rough estimates of the precision with
which the light horizon can be determined and some scalings.  We
then follow with a more detailed calculation, taking into
account possible degeneracies with parameters that affect the
B-mode power spectrum.

To begin with, consider an idealized full-sky (or nearly full sky) 
experiment and assume that the B-mode power spectrum is measured
with a detector-noise contribution $C_\ell^{\rm n}$; ignore for
now any lensing-induced \cite{Zaldarriaga:1998ar} B modes.
Consider the shift $C_\ell^{\rm BB} \to C_{\ell(1-\alpha)}^{\rm BB}$
in the B-mode power spectrum induced by a change $\delta r_{\rm
gw} = \alpha r_{\rm gw}$ in the light horizon.  We then estimate
the $1\sigma$ (68~\% C.L.) uncertainty with which the parameter
$\alpha$ can be determined, for an experiment that surveys a
fraction $f_{\rm sky}$ of the sky with noise power spectrum
$C_l^{\rm n}$ as
\begin{equation}
     \sigma_\alpha = \left[ \sum_\ell \frac{(2\ell+1) f_{\rm sky}}{2} 
         \left(
\frac{\partial C_\ell^{\rm BB}/\partial \alpha}{C_\ell^{\rm BB}+ C_\ell^{\rm n}}              
        \right)^2
 \right]^{-1/2}.
\label{eqn:sigmas}     
\end{equation}
The partial derivatives can be evaluated by 
$(\partial C_\ell^{\rm BB}/\partial \alpha) 
= - dC_\ell^{\rm BB}/d\ln \ell$.
For this estimation, we take the B-mode power
spectrum $C_\ell^{\rm BB}$ obtained for a scale-invariant
gravitational-wave power spectrum as expected from inflation.
Given that the signal we seek is the location of the peaks in
$C_\ell^{\rm BB}$, we take the reionization optical depth
$\tau=0$.  We take the sum from $\ell=20$ to $\ell=500$ (well within
the target range of a stage-IV CMB experiment; see
Fig.~\ref{fig:ClBB}).  We choose the lower limit, $\ell=20$ as
the recombination peak at $\ell\lesssim 10$ will not be shifted
by a change to the light horizon at the surface of last scatter.
In practice, the results are insensitive to changes in either
the lower or upper bounds, as the signal peaks near $\ell\sim100$.

We next determine $C_\ell^{\rm n}$ in terms of $\sigma_r$, the
smallest detectable (at $1\sigma$) tensor-to-scalar ratio, as it
is a commonly discussed figure of merit for B-mode searches.  We
thus estimate smallest detectable tensor-to-scalar ratio $r$ to be
\cite{Kamionkowski:1997av}
\begin{equation}
    \sigma_r = \left[ \sum_\ell
        \frac{(2\ell+1) f_{\rm sky}}{2} 
        \left(
     \frac{ \partial C_\ell^{\rm BB} /\partial r}{C_\ell^{\rm n}} 
 \right)^2
     \right]^{-1/2}.
\label{eqn:sigmar}     
\end{equation}
Note that the signal power spectrum $C_\ell^{\rm BB}$ does not
appear in the denominator here, as this expression is for the
error with which $r$ is measured under the null hypothesis
$r=0$.  We then use Eq.~(\ref{eqn:sigmar}) to
fix the noise power spectrum $C_\ell^{\rm n}$ in terms of
$\sigma_{\ln r} \equiv \sigma_r/r$, which has been carefully forecast in several
detailed studies of hypothetical or specific experimental
designs. In so doing, we circumvent issues involving
imperfectly subtracted foregrounds and lensing-induced B modes
(which act effectively as a contribution to $C_\ell^{\rm n}$) by
using results from these prior studies.

%%%%%%%%%%%%%%%%%%%%%%%%%%%%%%%%%%%%%%%%%%%%%%%%%%%%%%%%%%%%%%%%%%%%%%%%%%
\begin{figure}
    \includegraphics[width=0.49\textwidth]{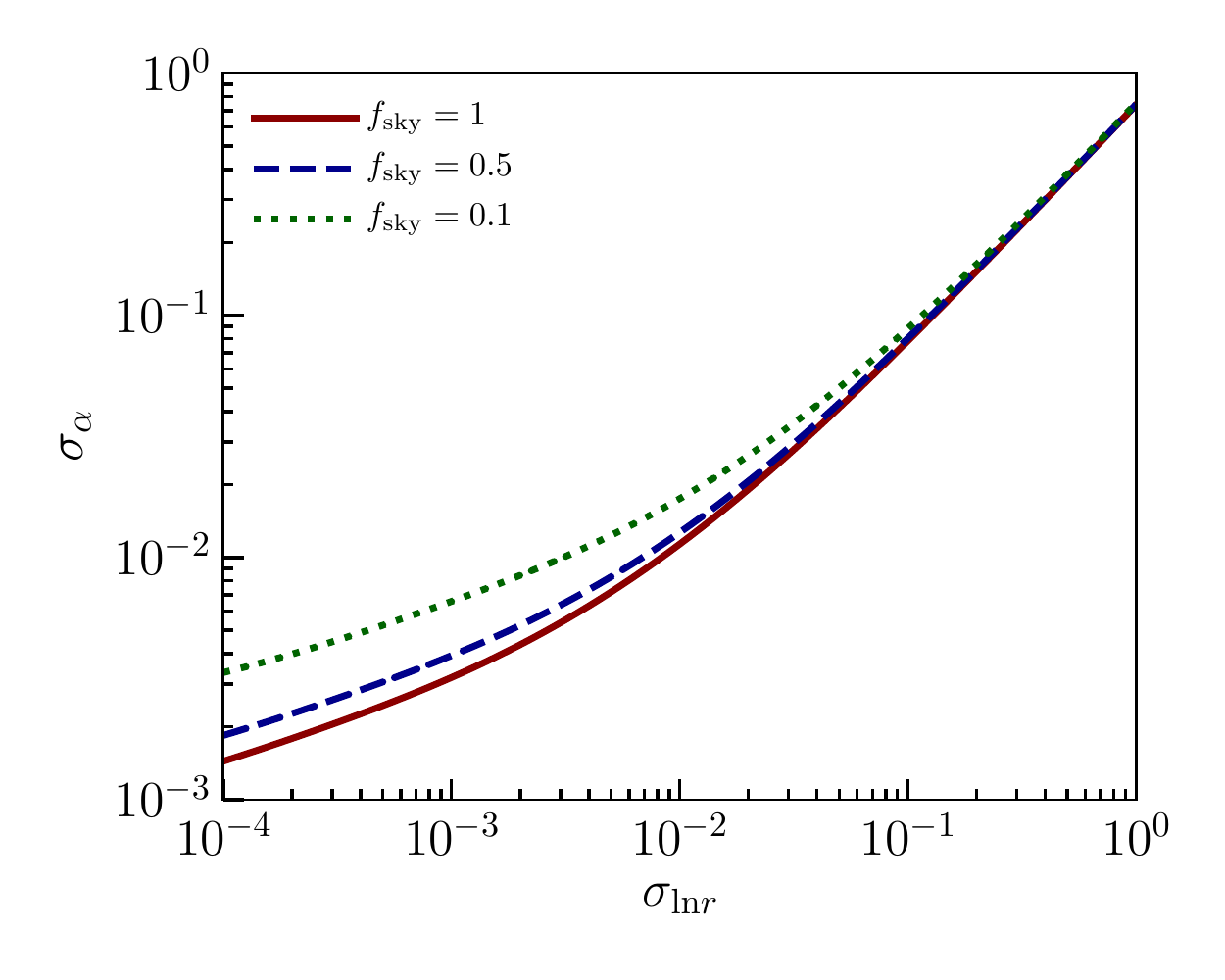}
\caption{
Projected 1-$\sigma$ (68~\% C.L.) accuracy of measuring the light horizon 
from the B-mode polarization power spectrum as a function of 
the inverse,  $\sigma_r/r =\sigma_{\ln r}$, of the
signal-to-noise ratio of the tensor-to-scalar
ratio. Here, we do not marginalize over the other parameters.
}
\label{fig:bestcase}
\end{figure}
%%%%%%%%%%%%%%%%%%%%%%%%%%%%%%%%%%%%%%%%%%%%%%%%%%%%%%%%%%%%%%%%%%%%%%%%%%

We show in Fig.~\ref{fig:bestcase} the error, inferred from
Eqs.~(\ref{eqn:sigmas}) and (\ref{eqn:sigmar}), with which
the B-mode peak location can be determined (at $1\sigma$) for
three different values of $f_{\rm sky}=1$ ({\it red solid} line), $0.5$
({\it blue dashed} line) and $0.1$ ({\it green dotted} line).  In the most optimistic
case that $r=0.06$, and with $\sigma_r=0.001$, the calculation
indicates a $\lesssim 2\%$ (at $1\sigma$) measurement of the
light-horizon distance at decoupling.  This is encouraging.

The numerical results in Fig.~\ref{fig:bestcase} are insensitive 
to the highest multipole moment $\ell_{\rm max}$ used in the sums in 
Eqs.~(\ref{eqn:sigmas}) and (\ref{eqn:sigmar}), and remain more or less 
the same for any $\ell_{\rm max}\gtrsim150$.  In other words, the meaurement
comes almost entirely from the first peak, the ``recombination
peak'' (which occurs at $l\simeq 86$), in the B-mode power
spectrum.  This also implies that the measurement requires the B
modes to be mapped with an angular resolution no better than
$1^\circ$.

%%%%%%%%%%%%%%%%%%%%%%%%%%%%%%%%%%%%%%%%%%%%%%%%%%%%%%%%%%%%%%%%%%%%%%%%%%
\begin{figure}
    \includegraphics[width=0.49\textwidth]{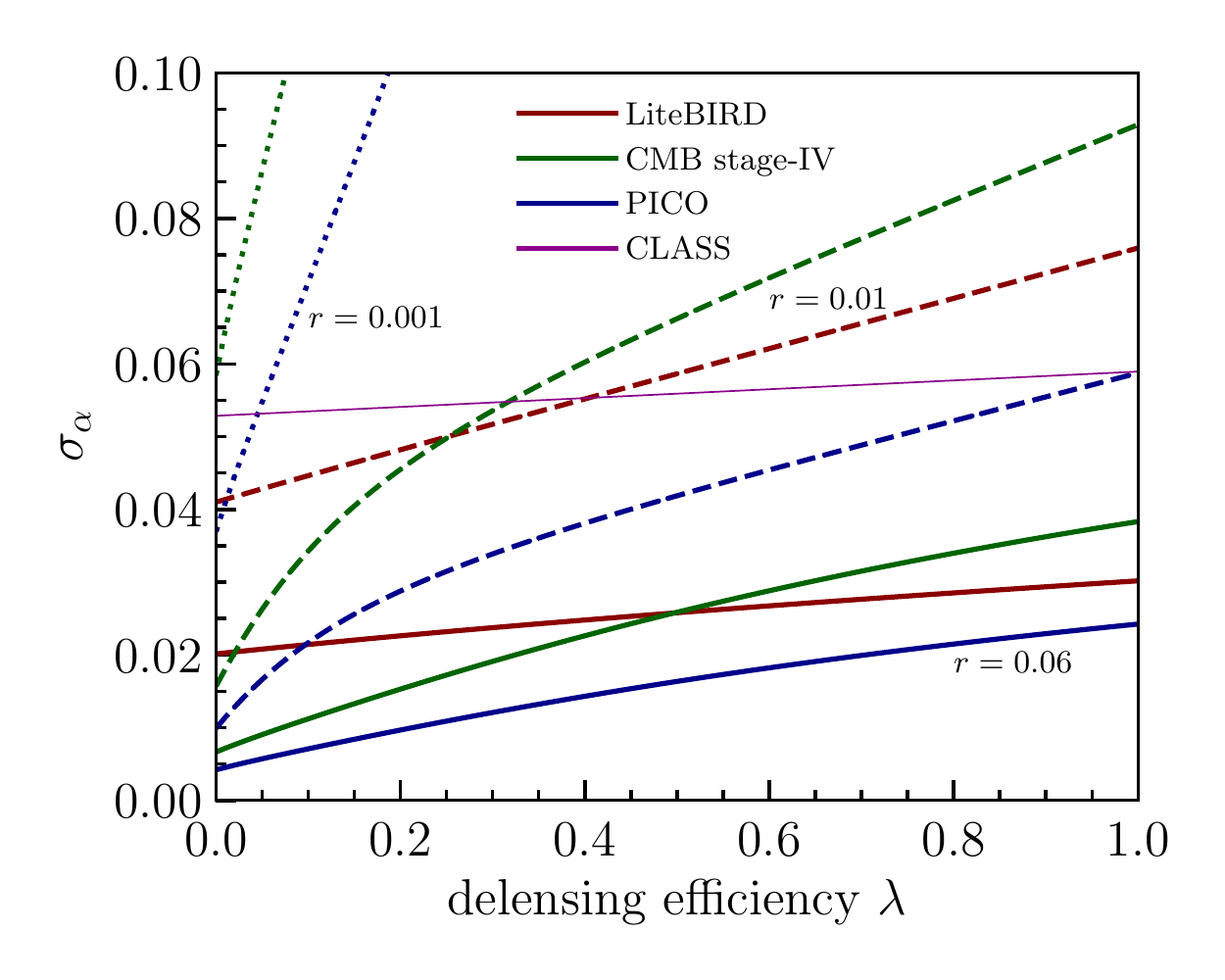}
\caption{
Marginzlized 1-$\sigma$ (68~\% C.L.) accuracy of measuring the light horizon 
from the B-mode polarization power spectrum from three future experiments:
LiteBIRD, CMB stage-IV and PICO, as a function of delensing efficiency 
$\lambda$ for $r=0.001$ ({\it dotted}), $r=0.01$ ({\it dashed}), 
$r=0.06$ ({\it solid}), after marginalizing over $r$, $n_t$, $\tau$ and 
$\lambda$. For comparison, we also show the same uncertainty for the 
CLASS mission when fixing $r=0.06$, $n_t=0$ with the thin magenta line.
}
\label{fig:Fisher}
\end{figure}
%%%%%%%%%%%%%%%%%%%%%%%%%%%%%%%%%%%%%%%%%%%%%%%%%%%%%%%%%%%%%%%%%%%%%%%%%%

We now follow up with a more careful Fisher forecast which takes
into account the possibility of imperfect subtraction of
lensing-induced B modes, possible shifts in the first-peak
location from uncertainties in the spectral index $n_t$ of the
gravitational-wave power spectrum, and covariances between the
different parameters.  We further include the effects of
reionization, as for some experiments (e.g., LiteBIRD), the
sensitivity to B modes may be dominated by the low-$\ell$
reionization peak, rather than the recombination peak assumed in
the simple estimates above.  We provide results for
experimental specifications that correspond roughly to those for
several projects being pursued or under consideration.
The Fisher matrix for the B-mode polarization power spectrum is
given as  \cite{Jungman:1995bz,Hiramatsu:2018nfa}
\begin{equation}
F_{ij}
=
\sum_{\ell}
\frac{f_{\rm sky}(2\ell+1)}{2}
\frac{1}{{\cal N}_\ell^2}
\left(\frac{\partial C_\ell^{\rm BB, obs}}{\partial\theta_i}\right)    
\left(\frac{\partial C_\ell^{\rm BB, obs}}{\partial\theta_j}\right)\,,    
\label{eq:Fisher}
\end{equation}
with $\theta=\{\alpha$, $r$, $n_t$, $\tau$, $\lambda\}$ the
vector of parameters being determined by the measurement of the
B-mode power spectrum.  Here $\tau$ is the reionization optical
depth and $\lambda\in[0,1]$ is the fraction of the
lensing-induced B modes that remain after de-lensing
\cite{Kesden:2002ku,Knox:2002pe}.

The observed power spectrum that appears in
Eq.~(\ref{eq:Fisher}) is then $C_{\ell}^{\rm BB, obs} = 
C_\ell^{\rm BB}(r,n_t,\tau) 
+ \lambda C_\ell^{\rm BB, lens}$, including a contribution from
imperfectly subtracted lensed-induced B modes.
We define the noise power per each harmonic mode as
${\cal N}_\ell = C_\ell^{\rm BB, obs}+C_\ell^{\rm n} \exp\left[\ell(\ell+1)\sigma_b^2\right]$, with the instrumental noise,
\begin{equation}
C_\ell^{\rm n} =
\left(
    \frac{\pi}{10800}
    \frac{w_p^{-1/2}}{\mu{\rm K}\,{\rm arcmin}}
\right)^2 \mu{\rm K}^2\,{\rm str}\,,
\end{equation}
and 
\begin{equation}
    \sigma_b = 
    \left(\frac{\pi}{180}\right)
    \frac{\theta_{\rm fwhm}}{\sqrt{8\ln2}}\,,
\end{equation}
with the full width at half maximum size $\theta_{\rm fwhm}$ of
the beam.  We adopt the following values for the
four types of experiments that we consider here:
$(w_p^{-1/2},\theta_{\rm fwhm},f_{\rm sky})=$
$(10 \mu{\rm K\,arcmin},60\,{\rm arcmin},0.4)$ for the CLASS mission,
$(3 \mu{\rm K\,arcmin},30\,{\rm arcmin},1)$ for the LiteBIRD satellite,
$(1 \mu{\rm K\,arcmin},3\,{\rm arcmin},0.4)$ for the ground-based 
CMB stage-IV experiments (Simons Observatory and CMB-S4), and
$(1 \mu{\rm K\,arcmin},3\,{\rm arcmin},1)$ for the PICO satellite.

For the fiducial cosmology, we use the best-fitting cosmological
parameters from Planck 2018 \cite{Aghanim:2018eyx} and calculate
primordial and lensing B-mode polarization power spectra by using
{\sf CAMB} \cite{Lewis:1999bs}.
We set $\partial C_\ell^{\rm BB,obs}/d\ln \ell=0$ for $\ell <
15$, as a shift in the light horizon in the early Universe will
not affect (by the model assumptions we are making here) the
light horizon at reionization.  Because $C_\ell^{BB}$ is almost
flat at large angular scales, the Fisher-matrix results are
insensitive to the exact value of $\ell$ we use for this cutoff.

As shown in the bottom panel of Fig.~\ref{fig:ClBB}, the derivative with
respect to the distance-scale (cyan line, $d\ln\ell$) has characteristic 
wiggles due to the acoustic peaks.  It turns out that the
light-horizon measurement is almost independent of the optical
depth ($\tau$) and lensing ($\lambda$), but moderately degenerate with
the amplitude ($r$) and slope ($n_t$) of the primordial
gravitational-wave power spectrum.  Note that within the context of
specific inflationary models, or classes of inflationary models,
further information on $n_t$ might be inferred from the
precise measurement of the scalar amplitude and spectral index
from the CMB temperature and E-mode power spectra.  If so, the
results we present here may err on the pessimistic side.

As expected, the results depend quite sensitively on $r$, and
de-lensing becomes increasingly important at lower values of $r$.
As shown in Fig.~\ref{fig:ClBB}, for $r=0.001$, the 
B-mode power spectrum is barely above the noise curve even for perfect
delensing ($\lambda=0$) and drops below the noise curve when using 
the moderate delensing efficiency ($\lambda=0.15$) expected from combining
various galaxy surveys \cite{Manzotti:2017oby}.
Indeed, we can see that in Fig.~\ref{fig:Fisher}, the projected 
uncertainties of measuring $\alpha=\delta r_{\rm gw}/r_{\rm gw}$ for 
$r=0.001$ (dotted lines) sharply rise beyond $\sigma_\alpha=10\%$ 
at $\lambda\simeq 0.1$.
There, we show the projected uncertainties on $\delta r_{\rm gw}/r_{\rm gw}$
for the three experiments (LiteBIRD, CMB stage-IV, PICO), after 
marginalizing over the other four parameters ($r$, $n_t$,
$\lambda$, $\tau$), as 
a function of the delensing efficiency $\lambda$.
As we have estimated earlier, for $r\lesssim0.06$ and
$\sigma_r\simeq 0.001$, we can measure the 
light-horizon scale to a few-percent level.  We have also verified
that for experiments like PICO and stage-IV, which target
primarily the recombination bump, the scalings in
Fig.~\ref{fig:bestcase} are valid.  The scalings are not
quite as effective, however, for an experiment like LiteBird
that targets primarily the low-$\ell$ reionization bump.

To summarize, a $\lesssim2\%$ determination of the angular scale
subtended by the light horizon at the
surface of last scatter is conceivable through measurement of
the B-mode power spectrum.  The CMB-polarization experiments are
similar to those being pursued already to detect the
B-mode signal, and could in the best-case scenario provide
results of the precision relevant for the Hubble tension on a
$\sim$decade timescale.  The measurement does require
that inflationary gravitational waves exist with a
tensor-to-scalar ratio $r$ not too much smaller than the current
upper bound, and
there is no way of telling, until the measurement is done,
whether Nature will cooperate in this regard.  If B modes are
detected and the peak location determined, it will narrow the
range of possible resolutions to the Hubble tension.  If the
result disagrees with the canonical expectation, it will rule
out late-time solutions to the Hubble tension.  If it agrees, it
will constrain (though not rule out categorically) early-time
solutions.

We also note, before closing, that the predictions assume that
gravitional waves propagate at the speed of light in the early
Universe.  This measurement can thus be used to test this
general-relativistic prediction at the $\sim2\%$ level, which
may be relevant for some alternative-gravity models.

\acknowledgments 
DJ was supported at Pennsylvania State 
University by NSF Grant No.\ AST-1517363 and NASA ATP Grant
No.\ 80NSSC18K1103, and 
MK was supported at Johns Hopkins in part by NASA Grant
no.\ NNX17AK38G, NSF Grant No.\ 1818899, and the Simons
Foundation.

%%%%%%%%%%%%%%%%%%%%%%%%%%%%%%%%%
%%%%%%%%%%% References %%%%%%%%%%%
%%%%%%%%%%%%%%%%%%%%%%%%%%%%%%%%%

\end{document}